\documentstyle[aps,prb,preprint,floats]{revtex}
\begin{document}
\draft
\title{\bf Collective Josephson Vortex Dynamics in\\
Long Josephson Junction Stacks}
\author{Ju H. Kim and J. Pokharel}
\address{
Department of Physics, University of North Dakota\\
P.O. Box 7129, Grand Forks, ND 58202-7129}
\maketitle

\begin{abstract}
We investigate the collective phase dynamics in conventional long
Josephson junction (LJJ) stacks and in layered superconductors,
exhibiting intrinsic LJJ behaviors.  Using a theoretical model which
accounts for both the magnetic induction effect and the breakdown of
local charge neutrality (i.e., charging effect), we show that the
collective motion of Josephson vortices, including the dispersion
of Josephson plasma mode and the Swihart-type velocity, in an 
intrinsic LJJ stack such as Bi$_2$Sr$_2$CaCu$_2$O$_{8+y}$ (BSCCO) is
significantly modified from those in a conventional LJJ stack.  In
BSCCO, the strength of the charging effect $\alpha$ is small (i.e.,
$\alpha \sim 0.1-0.4$), but it leads to notable changes in collective
phase dynamics, including changes to the stability condition.  Also,
we show that splitting of the supercurrent branch in the resistive
state is due to collective motion of Josephson vortices.  The width
of spread of these sub-branches in the linear current-voltage regime
depends on $\alpha$, suggesting another way to measure the charging
effect in BSCCO. 
\end{abstract}
\vskip 5mm


\vskip 5mm

\narrowtext

\section{Introduction}

Dynamics of magnetic vortices in a stack of long Josephson junctions
(LJJ) in a magnetic field applied parallel to the junction layers
have attracted much attention due to their intriguing applied and 
fundamental interests.  The motion of Josephson vortices in a single
junction system has been exploited in various devices.\cite{dev}
Collective motion of these vortices in a LJJ stack, in which layers
of superconductor (S) and insulator (I) are arranged vertically as 
in Fig. 1, can be exploited in high frequency devices\cite{hitc} such
as tunable submillimeter-wave oscillators and detectors.  Here
collective motion, including both {\it in-phase} and {\it out-of-phase}
modes shown in Fig. 2, arises from mutual phase-locking of Josephson
junctions caused by (magnetic) inductive coupling between screening
currents flowing around adjacent Josephson vortices as they move
under a bias current.  The phase-locking establishes phase coherence
across the Josephson junctions.  A LJJ stack exhibiting this phase
coherence leads to high power output and bandwidth, and it can serve
as a model system for scientific studies.\cite{Kim1,cheren}

The motion of Josephson vortices in LJJ stacks yields interesting
phenomena: (i) Josephson plasma resonance\cite{Mat,Kak} (JPR) and (ii) 
supercurrent sub-branching.\cite{UK,JULee,HJLee}  The experiments on
both conventional LJJ stacks\cite{UK} (e.g., Nb-Al/AlO$_x$-Nb
multilayers) and layered superconductors\cite{Mat,Kak,JULee,HJLee}
(e.g., Bi$_2$Sr$_2$CaCu$_2$O$_{8+y}$ (BSCCO)) behaving as intrinsic
LJJ stacks\cite{ijj} indicate that JPR can be tuned\cite{Bul0} by
magnetic field $B$ and can occur over a broad range of frequencies,
from microwave to submillimeter-wave.  Also, the supercurrent
branch in the current-voltage (I-V) data splits into multiple 
sub-branches when a bias current exceeds some critical value.\cite{JULee}
To explain the data,\cite{Mat,Kak,UK,JULee,HJLee} two theoretical
models have been proposed: one is based on the inductive coupling
(i.e., magnetic induction model),\cite{SBP} and the other is based on
the coupling due to charge variation in the S layers (i.e., charging
effect model).\cite{KT,chim}

The magnetic induction model assumes that the S layer thickness
$d_S$ is much larger than the Debye (charge screening) length
$r_D$ (i.e., $d_S \gg r_D$), as in conventional LJJ.  In this case,
charge variations (or electric field) at each S layers are screen
out, yielding local charge neutrality.  Consequently, the electric
field does not lead to the longitudinal coupling between the S
layers.  In this model, an applied magnetic field induces
supercurrents along the S layers and results in the inductive
interaction\cite{Efe,Vol,Bul} between adjacent S layers.  The
induction coupling strength is inversely proportional to the common
S layer thickness.  This model has been used to explain the 
experimental data for BSCCO.  However, the underlying assumption is
not justified in BSCCO since $d_S \sim 3\AA$ and 
$r_D \sim 2-3\AA$.\cite{KT,BSCCO}

On the other hand, the charging effect model accounts for the
nonequilibrium effect in atomic scale thick superconducting layers.
When the S layers are so thin to be comparable to the Debye length
(i.e., $d_S \sim r_D$), as in BSCCO, the breakdown of local charge 
neutrality yields the charging effect.\cite{KT}  The particle-hole
imbalance\cite{chim} may also occur since each superconducting
layers cannot completely screen out the charge variation.  Hence the
presence of charge variations yields the interaction between the
contiguous superconducting layers and leads to the coupling between
the S layers.  Recently the charging effect model, neglecting the
magnetic induction effect, has been used to interpret the data for
BSCCO.\cite{KT}

Earlier studies,\cite{SBP,KT} including numerical
simulations\cite{NM,NC} of a finite LJJ system, show that these two
models can explain the data qualitatively, but considerable
inconsistencies between the experimental and the theoretical results
have been found.  For example, transverse and longitudinal JPR
are predicted by the magnetic induction model and the charging
effect model, respectively, but the data indicate that both types of
resonance occur.\cite{LoTr}  Recent experiments\cite{JULee,HJLee} on
HgI$_2$-intercalated BSCCO and BSCCO single crystals indicate that
the supercurrent branch in the I-V data splits into multiple
sub-branches in the resistive state when $B \sim H_o$ (i.e., low
vortex density regime).  An estimated value\cite{Ho} of $H_o$ for
Nb-Al/AlO$_x$-Nb multilayers and BSCCO is roughly 0.001T and 0.2T, 
respectively.  In the dense vortex regime (i.e., $B \gg H_o$), the 
I-V data exhibit characteristic kinks, and these kinks closely
resemble the prediction made by Machida et al., using the magnetic
induction model.\cite{Machi}  However a closer examination\cite{HJLee}
of the data reveals some inconsistencies.\cite{HJLee}  These
inconsistencies suggest that a better theoretical model is needed to 
describe the LJJ stacks.

In this paper, we investigate collective phase dynamics in
conventional LJJ stacks and layered superconductors at low magnetic
fields (i.e., $B \sim H_o$ in which Josephson vortices are in every
I layers as in Fig. 2) and at low temperatures (i.e., below the 
Abrikosov vortex lattice melting temperature), using a theoretical
model accounting for both the induction effect and the charging
effect.  These two effects are equally important\cite{Kim2,Machi2}
in BSCCO since $r_D \sim d_S$, but the charging effect is neglected
in many studies because its strength $\alpha$ is small\cite{He}
(e.g., $\alpha \sim 0.1-0.4$ in BSCCO).  We show how the collective
motion of Josephson vortices is modified by a weak charging effect.
We outline two main results.  First, {\it the Josephson plasma
dispersion relation, the Swihart velocity, and the stability
condition for collective motion in BSCCO are considerably modified
from those in Nb-Al/AlO$_x$-Nb multilayers}.  Second, {\it the 
splitting of the supercurrent branch in the resistive state is due 
to collective motion of Josephson vortices}, and the width of 
spread of these sub-branches in the linear I-V regime depends on
$\alpha$.  These results are consistent with the experimental data
described above.

The remainder of the paper is organized as follows.  In Sec. II,
a theoretical model, which accounts for both the magnetic induction
effect and the charging effect, is derived by extending previous models.
In Sec. III, the Josephson plasma dispersion relation and the Swihart
velocity for the collective modes are computed from our model derived 
in Sec. II.  In Sec. IV, we determine the stability condition for the
mutually phase-locked modes, performing the linear stability analysis.
In Sec. V, we show that the splitting of the supercurrent branch in the
resistive state is due to the collective motion of Josephson vortices.
Finally, in Sec. VI, we summarize our results and conclude.  

\section {Theoretical model}

In this section, we derive a theoretical model, extending previous
approaches.  A brief discussion of this model was published.\cite{Kim2}
Here we consider a system with a large number of LJJ (i.e., $N \gg 1$)
neglecting the boundary effect and present new results obtained from 
this model in later sections.  To account for both the magnetic
induction effect and the charging effect, we start with the
gauge-invariant phase difference between the S layers $\ell$ and
$\ell-1$,
\begin{equation}
\varphi_{\ell,\ell-1} = \theta_\ell - \theta_{\ell-1} -
{2\pi \over \phi_o} \int_{\ell-1}^\ell {\bf A} \cdot d{\bf l}~,
\label{phase}
\end{equation}
where $\theta_\ell$ is the phase of the superconducting order
parameter, $\phi_o=hc/2e$ is the flux quantum, and $\bf A$ is the 
vector potential in the I layers.  In this paper, we employ the
Cartesian coordinates and assume that the S and I layers are stacked
along $z$-direction and the magnetic field is applied along the 
$y$-direction, as in Fig 1.  For simplicity, the thicknesses of the S 
($d_S$) and I ($d_I$) layers are taken to be uniform.  

The magnetic induction effect due to the applied magnetic field
(along the $y$-direction) yields a {\it spatial} variation of the
phase difference (along the $x$-direction).  An equation describing
the magnetic inductive coupling\cite{SBP} between the $S$ layers
\begin{equation}
{\phi_o \over 2 \pi}{\partial \varphi_{\ell,\ell-1} \over
\partial x} = s (B_{\ell+1,\ell} + B_{\ell-1,\ell-2} ) +
 d' B_{\ell,\ell-1}
\label{induct}
\end{equation}
is easily obtained by taking a {\it spatial} derivative of Eq.
(\ref{phase}) and by using the expression for the supercurrent
density
\begin{equation}
{\bf J}_\ell~=~ {\phi_o  \over 8\pi^2 \lambda^2}~
\left(\nabla \theta_\ell -
{2\pi \over \phi_o} {\bf A}_\ell \right)~.
\end{equation}
Here $d'=d_I+2\lambda\coth(d_S/\lambda)$ and
$s=-\lambda[\sinh(d_S/\lambda)]^{-1}$ are expressed in terms of
the London penetration depth $\lambda$.  The magnetic field
$B_{\ell,\ell-1}$ in the I layer between two S layers $\ell$ 
and $\ell-1$ is parallel to the layers.  Note that $B_{\ell,\ell-1}$
differs from $B$ since the magnetic field generated by the 
supercurrent in the S layers modifies the field in the I layer. 
Using Maxwell's equation, we express the spatial derivative of the
magnetic field as
\begin{equation}
{\partial B_{\ell,\ell-1} \over \partial x}~=~ {4\pi \over c}
(J_c\sin\varphi_{\ell,\ell-1}-J_B+J^T_{\ell,\ell-1})
\label{Max}
\end{equation}
where $J_c$ is the Josephson critical current density, and $J_B$ is
a bias current density.  Note that the magnetic field entering the
I layers yields\cite{BC} a triangular Josephson vortex lattice (JVL)
when the bias current is either absent or small.  The current
density\cite{chim} $J^T$, 
\begin{equation}
J^T_{\ell,\ell-1}~=~{\phi_o \over 2\pi}{\sigma \over {\cal D}}~
{\partial \varphi_{\ell,\ell-1} \over \partial t}~+~
{\epsilon \over 4\pi {\cal D}}~{\partial V_{\ell,\ell-1} \over
\partial t}~,
\label{curden}
\end{equation}
includes the quasiparticle and the displacement current contribution.
Here ${\cal D}=d'+2s=d_I+2\lambda \tanh (d_S/2\lambda)$ is the 
effective thickness of the block layer, $\sigma$ is the quasiparticle
conductivity, $\epsilon$ is the dielectric constant of I layers, and
$V_{\ell,\ell-1}$ is the voltage between the S layers $\ell$ and
$\ell-1$.  Using Eq. (\ref{Max}), we rewrite Eq. (\ref{induct}) as 
\begin{eqnarray}
{\phi_o \over 8\pi^2}{\partial^2\varphi_{\ell,\ell-1} \over
\partial x^2}
&=& J_c~\bigg [s(\sin\varphi_{\ell+1,\ell} +
\sin\varphi_{\ell-1,\ell-2}) + d' \sin\varphi_{\ell,\ell-1}
\bigg] - {\cal D}J_B
\nonumber \\
&+& d' J^T_{\ell,\ell-1} + s (J^T_{\ell+1,\ell} + 
J^T_{\ell-1,\ell-2})~.
\label{ic}
\end{eqnarray}
Note that ${\cal S}=s/d'$ measures the induction coupling strength,
and ${\cal S}=-0.5$ in the strong coupling limit.  The phase
difference equation (Eq. (4) in Ref. 27) derived by Bulaevskii and
Clem within the framework of Lawrence-Doniach model\cite{LD} can be
obtained from Eq. (\ref{ic}) when the time-dependent terms are
neglected (i.e., $J^T_{\ell,\ell-1}=0$) and the relations
$(8\pi^2/\phi_o)J_c d'=(2/\lambda_J^2)+(1/\lambda_c^2)$ and
$(8\pi^2/\phi_o)J_c s= -1/\lambda_J^2$ are used.  Here $\lambda_c$ 
is the magnetic penetration depth in the direction perpendicular to
the S layers.

The presence of a nonequilibrium state leads to the interaction
between the S layers.  When the S layer thickness is comparable to 
the Debye screening length (i.e., $r_D \sim d_S$), the S layers are
in a nonequilibrium state because the charge variations in these
layers are not completely screened.  This incomplete charge
screening enhances the {\it temporal} variation of the phase
difference.  One can include this effect in the phase dynamics,
modifying\cite{KT,chim} the usual AC Josephson relation, which is 
a {\it time} derivative of Eq. (\ref{phase}), to
\begin{equation}
{\phi_o \over 2\pi}{\partial \varphi_{\ell,\ell-1} \over
\partial t} = V_{\ell,\ell-1} + \Phi_\ell - \Phi_{\ell-1} 
\label{mjr}
\end{equation} 
as a way to account for a nonzero gauge-invariant potential
$\Phi_\ell=\phi_\ell+(\hbar/2e)(\partial\theta_\ell/\partial t)$
generated inside the S layers.  Here $\phi_\ell$ is the
electrostatic potential.  The modified AC Josephson relation of
Eq. (\ref{mjr}) can be rewritten as
\begin{equation}
{\phi_o \over 2\pi}{\partial \varphi_{\ell,\ell-1} \over \partial t}
= V_{\ell,\ell-1} - \alpha~(V_{\ell+1,\ell}-2 V_{\ell,\ell-1} +
V_{\ell-1,\ell-2}) - \Psi_\ell + \Psi_{\ell-1} ~,
\label{mjr1}
\end{equation}
using the charge density $\rho_\ell=-(\Phi_\ell-\Psi_\ell)/4\pi
r_D^2$ and the Maxwell's equation $\epsilon\nabla\cdot {\bf E} =
4\pi\rho$.  Here $\alpha=\epsilon r_D^2/{\cal D}d_S$ measures the
strength of the charging effect, and $\Psi_\ell$ measures the 
particle-hole imbalance in the S layer.  For simplicity, we consider
only the charging effect by setting $\Psi_\ell=\Psi_{\ell-1}=0$, as
it has been done in earlier studies.\cite{KT}  Note that the usual
AC Josephson relation is obtained from Eq. (\ref{mjr1}) when
$\alpha=0$.  This indicates that the charging effect (i.e.,
$\alpha \ne 0$) enhances the coupling between neighboring junctions.
Using Eq. (\ref{curden}), we relate\cite{KT} the time derivative of
the phase difference to the current densities and obtain 
\begin{eqnarray}
J_c~\left[ {1 \over \omega^2_p}
{\partial^2 \varphi_{\ell,\ell-1} \over \partial t^2} +
{\beta \over \omega_p}{\partial\varphi_{\ell,\ell-1}
\over \partial t} - {\alpha\beta \over \omega_p} {\partial 
\over\partial t} \bigg( \varphi_{\ell+1,\ell} - 2 
\varphi_{\ell,\ell-1} + \varphi_{\ell-1,\ell-2} \bigg) \right] =
\nonumber \\
J^T_{\ell,\ell-1} - \alpha~(J^T_{\ell+1,\ell} - 2 J^T_{\ell,\ell-1}
+ J^T_{\ell-1,\ell-2})~.~~~~~~~~~~~~~~~~~
\label{mjr2p}
\end{eqnarray}
Here $\omega_p=c/(\sqrt{\epsilon}\lambda_c)$ is the plasma frequency,
$\beta=(4\pi/c) (\sigma\lambda_c/\sqrt{\epsilon})=1/\sqrt{\beta_c}$, 
and $\beta_c$ is the McCumber parameter.  Note that the spatial
variation of $\varphi_{\ell,\ell-1}$ is neglected in the charging
effect model\cite{KT} of Eq. (\ref{mjr2p}).  The terms of the order
${\cal O}(\alpha\beta)$ can be safely neglected since $\alpha$ and
$\beta$ are small in the layered superconductors (i.e., $\alpha\beta
\ll 1$).  For example, the experimental value for $\alpha$ and $\beta$
in BSCCO are roughly 0.1-0.4 and 0.2, respectively.\cite{BSCCO,He}
Neglecting these small terms, we rewrite Eq. (\ref{mjr2p}) as
\begin{equation}
J_c \left( {1 \over \omega^2_p}
{\partial^2 \varphi_{\ell,\ell-1} \over \partial t^2} + 
{\beta \over \omega_p}
{\partial\varphi_{\ell,\ell-1} \over\partial t} \right) \approx 
J^T_{\ell,\ell-1} - \alpha~(J^T_{\ell+1,\ell} - 
2 J^T_{\ell,\ell-1} + J^T_{\ell-1,\ell-2}) ~.
\label{mjr2}
\end{equation}
As we shall see in Sec. III, the charging effect terms in Eq.
(\ref{mjr2}) yield purely longitudinal Josephson plasma
excitations.\cite{KT}

A theoretical model, including both the magnetic induction effect
and the charging effect, can be obtained easily by noting that the 
magnetic induction model of Eq. (\ref{ic}) and the charging effect
model of Eq. (\ref{mjr2}) are coupled to each other via the 
current density $J^T$.  Combining Eqs. (\ref{ic}) and (\ref{mjr2}),
we obtain the coupled sine-Gordon equations,
\begin{eqnarray}
{\partial^2\varphi_{\ell,\ell-1} \over \partial x^2} -
{1\over \lambda^2_c {\cal D}} \left( {1 \over \omega^2_p}
{\partial^2 \Delta_\ell \over \partial t^2} + 
{\beta \over \omega_p} {\partial \Delta_\ell \over\partial t}
\right)-{\alpha \over \lambda^2_c {\cal D}} {1 \over \omega^2_p}
{\partial^2 \Xi_\ell \over \partial t^2} ~=~
\nonumber \\
{1\over \lambda^2_c {\cal D}}
~\bigg [d' \sin\varphi_{\ell,\ell-1} +
s(\sin\varphi_{\ell+1,\ell}+\sin\varphi_{\ell-1,\ell-2}) -
{\cal D}{J_B \over J_c} \bigg ] ~,
\label{mf}
\end{eqnarray}
where $\Delta_\ell = d' \varphi_{\ell,\ell-1} +
s (\varphi_{\ell+1,\ell} + \varphi_{\ell-1,\ell-2})$ and 
$\Xi_\ell = s(\varphi_{\ell-2,\ell-3}+
\varphi_{\ell+2,\ell+1}) + (d'-2s)(\varphi_{\ell-1,\ell-2}+
\varphi_{\ell+1,\ell}) +2(s-d')\varphi_{\ell,\ell-1}$.  The third 
term on the left hand side of Eq. (\ref{mf}) (due to the charging 
effect) is the main modification from the earlier models.  Hence,
Eq. (\ref{mf}) becomes identical to the phase difference equation
derived by Bulaevskii et al. (Eq. (11) in Ref. 17) when $\alpha=0$.
Using Eq. (\ref{mf}), we show below that a weak charging effect in
BSCCO (i.e., $\alpha \sim 0.1-0.4$) can yield significant changes
to the phase dynamics. 

\section {Josephson plasma dispersion relation}

We now determine the dispersion relation for the Josephson plasma and 
the Swihart velocity for the collective modes, using linear analysis:
$\varphi_{\ell,\ell-1}=\varphi^{(0)}_{\ell,\ell-1} +
\varphi'_{\ell,\ell-1}$.  Here $\varphi'_{\ell,\ell-1}$ describes small
fluctuations about $\varphi^{(0)}_{\ell,\ell-1}$ describing uniform
motion of Josephson vortices in the I layer between $\ell$ th and
$\ell-1$ th S layers.  $\varphi^{(0)}_{\ell,\ell-1}$ is zero in the
Meissner state, but in general, it depends on a magnetic 
field,\cite{phch} allowing JPR to be tunned by the field.  The effect
of magnetic field on JPR can be accounted for more accurately via the
field dependence of $J_c$ and via imposing the boundary condition,
$(\phi_o/2\pi) (\partial\varphi_{\ell,\ell-1}/\partial x)={\cal D}B$,
explicitly at both $x=0$ and $x=L_x$ (a junction length) in numerical
simulations. 

When the bias current $J_B$ equals the Josephson current in each of
the I layers (i.e., $J_B = J_c \sin\varphi^{(0)}_{\ell,\ell-1}$), we
describe the motion of vortices in terms of a uniform motion
$\varphi^{(0)}_{\ell,\ell-1}$ and small perturbation
$\varphi'_{\ell,\ell-1}$ about $\varphi^{(0)}_{\ell,\ell-1}$.  The
uniform phase motion is described by
\begin{equation}
{\partial^2\varphi^{(0)}_{\ell,\ell-1} \over \partial x^2} -
{1\over \lambda^2_c {\cal D}} \left( {1 \over \omega^2_p}
{\partial^2 \Delta^{(0)}_\ell \over \partial t^2} + 
{\beta \over \omega_p} {\partial \Delta^{(0)}_\ell \over\partial t}
\right)-{\alpha \over \lambda^2_c {\cal D}} {1 \over \omega^2_p}
{\partial^2 \Xi^{(0)}_\ell \over \partial t^2}~=~0~,
\end{equation} 
while small fluctuations (i.e., $\varphi'_{\ell,\ell-1}$) about
$\varphi^{(0)}_{\ell,\ell-1}$ are described by
\begin{eqnarray}
{\phi_o \over 8\pi^2}{\partial^2\varphi'_{\ell,\ell-1} \over
\partial x^2}
=J_c~[d'\varphi'_{\ell,\ell-1}\cos\varphi^{(0)}_{\ell,\ell-1}
+ s(\varphi'_{\ell+1,\ell}\cos\varphi^{(0)}_{\ell+1,\ell} +
\varphi'_{\ell-1,\ell-2}\cos\varphi^{(0)}_{\ell-1,\ell-2})]
\nonumber \\
+ d' J^T_{\ell,\ell-1} + s (J^T_{\ell+1,\ell} +
J^T_{\ell-1,\ell-2})~,
~~~~~~~~~~~~~~~~
\label{smo1} \\
J_c~\left( {1 \over \omega^2_p}
{\partial^2 \varphi'_{\ell,\ell-1} \over \partial t^2} + 
{\beta \over \omega_p}
{\partial\varphi'_{\ell,\ell-1} \over\partial t} \right) = 
J^T_{\ell,\ell-1}- \alpha~(J^T_{\ell+1,\ell}-
2J^T_{\ell,\ell-1} + J^T_{\ell-1,\ell-2})~.~~~
\label{smo2}
\end{eqnarray}
Equations (\ref{smo1}) and (\ref{smo2}) are coupled through the
current density $J^T_{\ell,\ell-1}$, suggesting that these 
equations can be simplified by expressing $\varphi'_{\ell,\ell-1}$
and $J^T_{\ell,\ell-1}$ as Fourier series in space for the 
$z$-direction: $\varphi'_{\ell,\ell-1}=\sum_{m=1}^{2N+1}T_m
e^{ik_m\ell a}$ and $J^T_{\ell,\ell-1}=\sum_{m=1}^{2N+1} J^T_m
e^{ik_m \ell a}$. $k_m=m\pi/(N+1)a$ represents the wavenumber for
the collective mode along the $z$-direction, $a=d_I+d_S$, $m$ is
the mode index, and $N$ represents the number of Josephson
junctions in a stack.

The Josephson plasma mode dispersion relation is determined easily by
approximating that $\varphi_{\ell,\ell-1}^{(0)} \approx \varphi^{(0)}$
and by combining Eqs. (\ref{smo1}) and (\ref{smo2}) into a single
equation as
\begin{equation}
{\partial^2 T_m \over \partial t^2}+
\omega_p^2 ~\bigg [ (k_x\lambda_c)^2 {\cal A}_m {\cal B}_m + 
{\cal A}_m \cos\varphi^{(0)} \bigg ]~T_m \approx 0~,
\label{matho}
\end{equation}
where ${\cal A}_m=1+4\alpha \sin^2(k_ma/2)$, ${\cal B}_m=\lbrace 
1+4[-{\cal S}/(1+2{\cal S})] \sin^2 (k_ma/2) \rbrace^{-1}$, and
${\cal S}=s/d'$.  Here, we set $\beta=0$ for simplicity.  From Eq.
(\ref{matho}), we obtain the dispersion relation of
\begin{equation}
\omega(k_x,k_m)= \omega_p~
\left[ (k_x\lambda_c)^2 {\cal A}_m {\cal B}_m +
{\cal A}_m \langle \cos \varphi^{(0)} \rangle_t \right]^{1/2}
\label{disp}
\end{equation} 
for the collective mode.  $\langle \cdot\cdot\cdot \rangle_t$ 
represents thermal averages.  The dispersion relation of Eq. 
(\ref{disp}) naturally recovers both purely longitudinal\cite{KT}
and purely transverse plasma excitations\cite{SBP} at $k_x=0$ and
at $k_m=0$, respectively.  However, there are notable differences
between our result of Eq. (\ref{disp}) and the results from other
models.\cite{SBP,KT}  Figure 3 illustrates the difference between
the dispersion relation of our model and that of the magnetic
induction model\cite{SBP} (Fig. 3(a)) and that of the charging 
effect model\cite{KT} (Fig. 3(b)).  

The changes in the dispersion relation due to the charging effect
increase the characteristic velocity of the collective mode.  The
group velocity for the electromagnetic waves in these LJJ is 
easily determined from Eq. (\ref{disp}) by evaluating
\begin{equation}
{d \omega \over dk_x}~=~{\omega_p^2 \over \omega}~
k_x\lambda_c^2~{\cal A}_m~{\cal B}_m~~
\end{equation}
within the linearized model.  This group velocity, asymptotically 
(i.e., as $k_x \rightarrow \infty$), leads to the Swihart velocity
\begin{equation}
{\bar c}_m~=~c_o \left[ {1+4\alpha\sin^2 (k_ma/2) \over
1+4 \left({-{\cal S} \over 1+2{\cal S}}\right) \sin^2 (k_ma/2)}
\right]^{1/2}~,
\label{Swi}
\end{equation}
the effective maximum velocity for the collective mode $m$.  Here
$c_o=c/\sqrt{\epsilon}$.  Equation (\ref{Swi}) recovers the result 
of the magnetic induction model (${\bar c}_m^{MI}$) when $\alpha =0$
(i.e., ${\bar c}_m^{MI}={\bar c}_m(\alpha=0)$), indicating that the
charging effect yields the mode-dependent enhancement of the Swihart
velocity from ${\bar c}_m^{MI}$.  For example, the Swihart velocity
is not enhanced for the $m=1$ mode (i.e.,
${\bar c}_1={\bar c}_1^{MI}$), but it is enhanced for the $m=N$ mode
(i.e., ${\bar c}_N= (1+4\alpha)^{1/2}{\bar c}_N^{MI}$).  This
enhancement reflects the increase in the coupling strength between 
the junctions due to the charging effect and indicates that 
the threshold velocity $v_{th}$ (=${\bar c}_N$) for emitting
Cherenkov radiation\cite{cheren} (i.e., non-Josephson emission) is
also increased.  For example, $v_{th}={\bar c}_N^{MI}$ when
$\alpha=0$, but $v_{th}=1.34{\bar c}_N^{MI}$ when $\alpha=0.2$.  
Evidence, indicating the need to account for the charging effect,
may be also found in the I-V data for BSCCO.  Recent analysis of the
I-V data in the dense vortex regime (i.e., $B \gg H_o$) indicates
that a better agreement between the predicted and observed position 
of the kinks can be obtained if the Swihart velocity for $m>1$ is
slightly larger than ${\bar c}_m^{MI}$.\cite{HJLee}  This suggests
that accounting for the charging effect is important for
quantitative understanding of the kinks in the I-V curves.

In Fig. 4, we compare the Josephson plasma mode dispersion for (a) 
Nb-Al/AlO$_x$-Nb multilayers and (b) BSCCO in the Meissner state (i.e.,
$\langle \varphi^{(0)} \rangle_t \approx 0$), using of Eq. (\ref{disp}).
To illustrate the difference between the dispersion of collective mode
for these two systems, we use the experimental values for the
parameters $\alpha$ (i.e., charging effect strength) and ${\cal S}$ 
(i.e., induction coupling strength).  For the spectrum corresponding to
the Nb-Al/AlO$_x$-Nb multilayers (Fig. 4(a)), we chose $\alpha=0.0$ and
${\cal S} \sim -0.47$ (assuming $\lambda \sim 900 \AA$, $d_I \sim 20\AA$,
and $d_S \sim 30\AA$).\cite{multi}  Here we chose $\alpha=0$ since the
charging effect is negligible when $d_S$ is much larger than an atomic
length.  For the spectrum corresponding to BSCCO (Fig. 4(b)), we chose
$\alpha=0.2$ and ${\cal S}\sim -0.49999$ (since $\lambda\sim 1500\AA$,
$d_I\sim 15\AA$, and $d_S \sim 3 \AA$).\cite{BSCCO}  Here $\alpha=0.2$
is chosen.  There are two notable differences between Figs. 4(a) and
4(b).  First, due to a stronger inductive coupling (i.e.,
${\cal S}=-0.49999$ versus $-0.47$), the frequency $\omega/\omega_p$,
for a fixed $k_x\lambda_c$, near $k_ma=0$ in Fig. 4(b), decreases more 
sharply with $k_ma$ than that in Fig. 4(a).  Second, due to the
charging effect (i.e., $\alpha=0.2$ versus 0.0), the collective mode
frequency for $k_x\lambda_c=0$ shows a dispersion as a function of $k_m$
in Fig. 4(b), indicating purely longitudinal excitations, while no
dispersion is shown in Fig. 4(a), indicating the absence of these
excitations.  Note that the effect of finite, but small, $\beta$ is
negligible, here. 

\section{Stability of collective modes} 

In this section, we discuss the stability of uniform motion of 
collective modes (i.e., moving JVL) shown in Fig. 2 against small
fluctuations.  The structure of the moving JVLs, driven by a bias
current, evolves as a function of its velocities.\cite{stab}  This
evolution can be easily understood in terms of the stable-unstable
transition for the collective modes.  

We now carry out the linear analysis and determine the condition
for maintaining stable uniform motion (i.e., the condition for
bound oscillations of $\varphi'_{\ell,\ell-1}$) by computing the
velocities at which the driven collective modes are stable.  Here,
instability of uniform motion arises when the amplitude of
fluctuations grows exponentially as the collective modes propagate
along the junction layers.  Similar analysis, not including the
charging effect, have been carried out to investigate the stability
of moving JVL against lattice deformation.\cite{stab}  Also the
effects of quantum and thermal fluctuations have been 
studied.\cite{Kim1}  We note that accounting for either the dynamic
phase transition\cite{dynam} induced by the lattice displacements 
or the fluctuation effects are beyond the scope of the present
analysis. 

We proceed the analysis writing the spatial and the temporal
dependence of phase fluctuations (i.e., $\varphi'_{\ell,\ell-1}$) of
Eqs. (\ref{smo1}) and (\ref{smo2}) in Fourier space for the 
$z$-direction.  Combining Eqs. (\ref{smo1}) and (\ref{smo2}) in 
Fourier space, we obtain 
\begin{equation}
{\lambda_c^2 {\cal D} \over d'{\cal C}_m}{\partial^2 T_m \over
\partial x^2} -{{\cal A}_m \over \omega_p^2} 
{\partial^2 T_m \over \partial t^2} - \left(
{\beta \over \omega_p}{\partial T_m \over \partial t} +
\cos\varphi_m^{(0)} T_m \right) =0
\label{small}
\end{equation}
where ${\cal C}_m=1+2{\cal S}\cos k_ma$.  The uniform motion of
the phase locked mode $\varphi_m^{(0)}$ with the wavenumber $k_m$
is given by $\varphi_m^{(0)}={\bar \omega}_m t+k_x x+
\varphi_{o,m}$,\cite{KT,Wat} where ${\bar \omega}_m=
(2\pi/\phi_o) V_m {\cal A}_m$ is the Josephson frequency, $V_m $
is the average voltage, and $\varphi_{o,m}$ is a mode dependent
constant.\cite{stab}   Here, the induced field contribution to
$\varphi_m^{(0)}$ from the Josephson effect is neglected.  This
contribution is negligible when the magnetic vortices are in
every I layers, as the case for $B \sim H_o$.  ${\bar \omega}_m/k_x$ 
is the velocity of the collective mode $m$.  We transform Eq.
(\ref{small}) into a familiar Mathieu equation in the following two
steps: first, make a change of variables from $(x,t)$ to $\zeta_m
(=\varphi_m^{(0)})$; and second, let $T_m= {\bar T}_m
e^{-\Gamma_m \zeta_m/2}$.  Here $\Gamma_m=\beta{\bar\omega}_m\omega_p
{\cal C}_m/\Omega_m$ and $\Omega_m= {\cal A}_m{\cal C}_m \omega_m^2
-(k_x\lambda_c)^2 ({\cal D}/d')\omega_p^2$.  The stability condition
is determined, solving
\begin{equation}
{\partial^2 {\bar T}_m \over \partial \zeta_m^2} + [
{\bar \delta}_m^T + {\bar \eta}_m^T \cos \zeta_m ] {\bar T}_m =0~,
\label{mathp}
\end{equation}
where ${\bar\delta}_m^T=-\Gamma_m^2/4$ and ${\bar\eta}_m^T={\cal C}_m
\omega_p^2/\Omega_m$ are the mode dependent (i.e., $k_m$) parametric
constants.  Solutions of Eq. (\ref{mathp}) exhibit 
instability\cite{Wat,Pede} for certain values of ${\bar\delta}_m^T$
and ${\bar\eta}^T_m$, indicating that the collective mode becomes 
unstable against small fluctuations.  We determine the stability
condition, finding $({\bar\delta}_m^T,{\bar\eta}_m^T)$ at which all
solutions of Eq.  (\ref{mathp}) are bounded.  Note that a similar
parametric instability (in the ${\bar\delta}_m^T-{\bar\eta}^T_m$ space)
occurs both in the magnetic induction model\cite{Pede} (i.e.,
$\alpha=0$) and in the charging effect model\cite{KT} (i.e.,
$k_x\lambda_c=0$). 

For finding the stability condition for the collective modes, it is
useful to determine, first, the stability diagram of the Mathieu 
equation 
\begin{equation}
{d^2 T \over d\zeta^2} + [\delta + \eta \cos \zeta] T = 0 ~,
\label{math}
\end{equation} 
and then, find the values of $({\bar\delta}_m^T,{\bar\eta}_m^T)$ 
corresponding to the stable region of this diagram.  The boundary
curves separating the region of bound (stable) and unbound
(unstable) solutions can be obtained easily, solving Eq. (\ref{math})
numerically following the procedure outlined in Ref. 35.  The boundary
curves for the periodic oscillations with the period $2\pi$ (i.e.,
$\zeta=2\pi$) and $4\pi$ (i.e., $\zeta=4\pi$) are obtained, imposing
that the determinant ${\cal E}_n$ for $n=\infty$, derived from Eq.
(\ref{math}) writing $T=\sum_{n=-\infty}^{n=\infty} C_n e^{in\zeta}$
for $\zeta=2\pi$ and $T=\sum_{n=-\infty}^{n=\infty} d_n e^{in\zeta/2}$
for $\zeta=4\pi$, is zero (i.e., ${\cal E}_n=0$).\cite{JS}  Here 
${\cal E}_n$ is the determinant of a $(2n+1)\times (2n+1)$ matrix for
a periodic solution with the period $2\pi$ (or a $2n\times 2n$ matrix
for a periodic solution with the period $4\pi$).  The determinant
${\cal E}_n$ can be computed using the recursion relation\cite{JS}
\begin{equation}
{\cal E}_{n+2} = (1-\gamma_{n+2}\gamma_{n+1}) {\cal E}_{n+1} -
\gamma_{n+2}\gamma_{n+1}(1-\gamma_{n+2}\gamma_{n+1}) {\cal E}_n +
\gamma_{n+2}\gamma_{n+1}^3\gamma_n^2 {\cal E}_{n-1}
\label{recur}
\end{equation}
where ${\cal E}_0=1$, ${\cal E}_1=1-2\gamma_0\gamma_1$,
${\cal E}_2=(1-\gamma_1\gamma_2)^2-2\gamma_0\gamma_1(1-\gamma_1
\gamma_2)$, and $\gamma_n=\eta/[2(\delta - n^2)]$ for a
solution with the period-$2\pi$ and ${\cal E}_1=1-\gamma_1^2$, 
${\cal E}_2=(1-\gamma_1\gamma_2)^2-\gamma_1^2$, 
${\cal E}_3=(1-\gamma_1\gamma_2-\gamma_2\gamma_3)^2-
\gamma_1^2(1-\gamma_2\gamma_3)^2$, and $\gamma_n=2\eta/[4\delta - 
(2n-1)^2]$ for a solution with the period-$4\pi$.

The stability diagram for Eq. (\ref{math}) is shown in Fig. 5.  The
unstable regions, where at least one solution is unbounded, are 
shaded, and the stable regions, where all solutions are bounded, 
are not shaded.  The boundary curves separating these regions are
periodic solutions with period $2\pi$ (dashed lines) and $4\pi$
(solid lines).  These curves are obtained by calculating
${\cal E}_n=0$ for $n=200$. The filled squares represent the values
of $({\bar\delta}_m^T,{\bar\eta}_m^T)$ satisfying the stability
condition.   Here we set ${\bar\delta}_m^T=0$ (i.e., $\beta=0$)
since ${\bar\delta}_m^T \propto \beta^2$ and the terms of 
this order ${\cal O}(\beta^2)$ have been neglected due to small 
$\beta$.  For ${\bar\delta}_m^T=0$, the following values satisfy the
stability condition: the values shown in Fig. 5 are 
$0<{\bar\eta}_m^T<0.4540$, ${\bar\eta}_m^T\approx$3.7898, 10.6516,
and 20.9637, and the values not shown in Fig. 5 are ${\bar\eta}_m^T
\approx$34.7142, 51.9022, 72.52784, 96.5910, 124.0918
$\cdot\cdot\cdot$.  For a large ${\bar\eta}_m^T$ satisfying the
stability condition, we assume that $\Omega_m \approx 0$ since
$\Omega_m\rightarrow 0$ as ${\bar\eta}_m^T \rightarrow \infty$.  In
this case, the velocity for uniform motion is given by 
\begin{equation}
{{\bar \omega}_m \over k_x} \approx \lambda_c \omega_p \left(
1+2{\cal S} \over {\cal A}_m {\cal C}_m \right)^{1/2}~,
\label{stabvel}
\end{equation}
indicating that the presence of the charging effect yields the mode
dependent modification to the stability condition.  For example,
${\bar \omega_1}/k_x \approx \lambda_c \omega_p$ for $m=1$ (i.e.,
rectangular lattice) but ${\bar \omega}_N/k_x \approx \lambda_c
\omega_p [(1+2{\cal S})/(1-2{\cal S})(1+4\alpha)]^{1/2}$ for $m=N$
(i.e., triangular lattice).  The velocity for the {\it out-of-phase}
modes is reduced from the predicted value of the magnetic induction
model (i.e., $\alpha=0$).  Equation (\ref{stabvel}) indicates that
moving Josephson vortices in a periodic array evolve from one stable
mode to another as the vortex velocity increases.  For example, as
the vortex velocity exceeds ${\bar \omega}_N/k_x$, but less than 
${\bar \omega}_{N-1}/k_x$, the moving triangular lattice ($m=N$)
becomes unstable and the $m=N-1$ mode becomes stable.

\section{Multiple sub-branching of supercurrent} 

In the resistive state, the supercurrent branch splits into multiple
sub-branches as the bias current exceeds the Josephson 
current.\cite{UK,JULee,HJLee}  Note that these supercurrent
sub-branches differ from the observed multiple quasiparticle
branches\cite{SBP,NM,KT} in the I-V data for LJJ stacks.  This
supercurrent sub-branching phenomenon, which appears clearly in the
non-linear I-V regime, is attributed to the motion of Josephson
vortices, but its origin is not understood clearly.  Microwave induced
voltage steps\cite{Shap} and geometric resonance\cite{geo} are
considered as other mechanisms, but we do not discuss them here.
Instead, we argue that the splitting of the supercurrent branch is
indeed due to the collective motion of Josephson vortices examining 
the low bias current regime where the I-V characteristics is linear.  
An analytic calculation is more tractable in this regime.  Here, we
illustrate qualitatively, rather than quantitatively, how the 
charging effect modifies the supercurrent sub-branches since the 
particle-hole imbalance effect\cite{chim} neglected in this study may
also need to be included for a quantitative comparison with the I-V data. 

The current-voltage relations in the resistive state is obtained
easily by noting that an AC voltage ripple with the Josephson
frequency $\omega_{\ell,\ell-1}$, in addition to the DC voltage,
appears across the junction when a bias current ($J_B$), greater than
the critical current, is applied.  This AC voltage ripple is due to
the electron-pair tunneling current across the junction.  Using the
modified AC Josephson relation of Eq. (\ref{mjr1}), the time dependence
of the phase difference between $\ell$ th and $\ell-1$ th S layers can
be written as\cite{Orl}
\begin{equation}
\varphi_{\ell,\ell-1}(t) \approx \varphi_{\ell,\ell-1}(0) +
{\bar \omega}_{\ell,\ell-1} t + {\phi_o \over 2\pi}{V_{\ell,\ell-1}^s
\over {\bar \omega}_{\ell,\ell-1}} \sin \omega_{\ell,\ell-1} t
\label{phase1}
\end{equation}  
where ${\bar\omega}_{\ell,\ell-1} = (2\pi/\phi_o) [\langle
V_{\ell,\ell-1} \rangle + \alpha (\langle V_{\ell+1,\ell}\rangle -2
\langle V_{\ell,\ell-1} \rangle + \langle V_{\ell-1,\ell-2} \rangle)]$
is the Josephson frequency, $\langle V_{\ell,\ell-1} \rangle$ is the 
DC voltage (i.e., time averaged) across the superconductor layers $\ell$
and $\ell-1$.  $V_{\ell,\ell-1}^s$ is the amplitude of the AC voltage
ripple.  This time dependent phase difference of Eq. (\ref{phase1})
yields a DC critical current response\cite{Orl} of
\begin{equation}
J_c \sin \varphi_{\ell,\ell-1}(t) = - {\rm J}_1 \left(
{\phi_o \over 2\pi}{V_{\ell,\ell-1}^s \over 
{\bar \omega}_{\ell,\ell-1}} \right)~\sin \varphi_{\ell,\ell-1}(0)
\label{rescur}
\end{equation}
across the $\ell$ th and $\ell-1$ th $S$ layers, indicating that the
junction becomes resistive when $J_B$ exceeds the DC critical current.
Here ${\rm J}_1(x)$ is the first order Bessel function of the first
kind.  Equation (\ref{rescur}) indicates that the current
$J_{\ell,\ell-1}= J_B - J_c \sin \varphi_{\ell,\ell-1}(t)$ between 
two adjacent S layers is not uniform along $z$-direction, even though
a uniform bias current is applied.  Hence, in this resistive state,
we may reduce Eq. (\ref{mf}) to
\begin{equation}
\left( {1 \over \omega^2_p} {\partial^2 \Delta_\ell \over 
\partial t^2} + {\beta \over \omega_p} {\partial \Delta_\ell 
\over\partial t} \right) +
{\alpha \over \omega^2_p}{\partial^2 \Xi_\ell \over \partial t^2} =
{1 \over J_c} \bigg [d' J_{\ell,\ell-1} + s(J_{\ell+1,\ell} +
J_{\ell-1,\ell-2}) \bigg]~.
\label{mf3}
\end{equation}
Here, we neglected the spatial dependence (i.e., $x$ variation) of
$\varphi_{\ell,\ell-1}$ for simplicity.  To explicitly express the
I-V relation for each collective mode, we now rewrite Eq. (\ref{mf3})
in Fourier space for the $z$-direction as
\begin{equation}
{{\cal A}_m \over \omega^2_p}
{\partial^2 \varphi_m \over \partial t^2} +
{\beta \over \omega_p} {\partial \varphi_m \over\partial t} 
={J_m \over J_c}~.
\label{math5}
\end{equation}
Note that the first and the second term on the left hand side of Eq.
(\ref{math5}) represent the capacitive and the resistive contribution
of the junction, respectively.

We now average Eq. (\ref{math5}) over time.  Since AC Josephson
tunneling leads to a small voltage oscillation about the DC voltage,
a further simplification of Eq. (\ref{math5}) can be made.  The
modified AC Josephson relation of $\partial \varphi_m/\partial t
= (2\pi/\phi_o) V_m {\cal A}_m$ indicates that the capacitive
contribution vanishes when it is averaged over time (i.e., $\langle
\partial^2\varphi_m/\partial t^2 \rangle \propto \partial\langle V_m
\rangle/\partial t \approx 0$).  This simplification leads to the 
current-voltage relation of
\begin{equation}
\langle V_m \rangle = {\omega_p \phi_o \over 2\pi \beta}
{{\bar J} \over J_c} {1 \over {\cal A}_m}~.
\label{iv}
\end{equation}
Here $\langle \cdot\cdot\cdot \rangle$ denotes the time average, and 
${\bar J}=\langle J_m \rangle \sim J_B-J_c \langle \sin\varphi
\rangle$.  Since the collective modes for $m=1,2, \cdot\cdot\cdot N$
are identical to the modes for $m=2N+1,2N, \cdot\cdot\cdot N+2$,
respectively, the number of sub-branches is the same as the number of
junctions (i.e., $N$) in the stack. 

In Fig. 6, we plot the I-V relation of Eq. (\ref{iv}) for (a)
$\alpha=0.0$ and (b) 0.1 to illustrate the effect of weak, but 
non-zero, charging effect.  For clarity, we plot the curves for only 
the three collective modes (i.e., $m=1$, $N/2$, and $N$) corresponding
to the modes shown in Fig. 2.  These I-V curves reveal two interesting
points.  First, the supercurrent splits into $N$ sub-branches, each
corresponding to the collective mode, when the LJJ stack are in the
resistive state.  The $m=1$ mode represents the high velocity mode 
(i.e., rectangular lattice), while the $m=N$ mode represents the low 
velocity mode (i.e., triangular lattice).  Second, these $N$
sub-branches appear as a single curve when the charging effect is
absent (i.e., $\alpha=0$, see Fig. 6(a)), but they spread out when
this effect is present (i.e., $\alpha \not= 0$, see Fig. 6(b)),
suggesting that this can be used as another way to measure the charging
effect.  Since the width of this spread is related to the strength of 
the charging effect ($\alpha$), identification of each sub-branches is
feasible at a low bias current.  When $\alpha$ is small, as in BSCCO
(i.e., $\alpha \sim 0.1-0.4$), observing the branch splitting in the
linear I-V regime may be difficult but is still possible.  The main
difficulty is in observing the high velocity branches (i.e., $m \sim
{\cal O}(1)$).  To observe these branches, a magnetic field, stronger
than $B \sim H_o$, may be needed because of their stability conditions.
Note that the appearance of these high velocity branches is expected
when the interaction between the vortices is increased by the
field,\cite{JULee} suggesting that a complete sub-branch structure may
be more easily obtained from the I-V characteristics of a LJJ stack
with increasing microwave irradiation power (i.e., AC magnetic
fields).\cite{HJLee} These results are consistent with the
data\cite{UK,JULee,HJLee} exhibiting supercurrent branch splitting. 

\section{summary and conclusion}

In summary, we investigated the collective phase dynamics in the
conventional LJJ stacks and in layered superconductors, using a 
theoretical model which accounts for both the magnetic induction
effect and the charging effect.\cite{SBP,KT}  These two coupling
mechanisms are equally important in the intrinsic LJJ (e.g. BSCCO)
due to the atomic length thick  S layers.  We showed that the 
collective phase dynamics in an intrinsic LJJ stack is
modified from those in a conventional LJJ stack in two important
ways.  (i) The dispersion of Josephson plasma mode for BSCCO is
significantly changed from the Nb-Al/AlO$_x$-Nb multilayers. 
Consequently, the Swihart velocity and the velocity of stable
uniform motion for the {\it out-of-phase} collective modes in BSCCO
increases and decreases, respectively, from the results of the 
magnetic induction model due to the presence of the charging effect.
(ii) The supercurrent sub-branching in the resistive state is
consistent with collective motion of Josephson vortices.  The width
of spread of these supercurrent sub-branches in the linear I-V 
regime depends on the strength of the charging effect.  These 
results are consistent with the experimental data and illustrate the 
importance of accounting for the charging effect in BSCCO, even though
its strength is weak ($\alpha \sim 0.1 - 0.4$).  They also suggest
that our model is useful for understanding the experimental data for
JPR, non-Josephson emission, and the I-V characteristics in the 
resistive state.  Since many applications of intrinsic LJJ stacks as 
high frequency devices exploit collective dynamics of Josephson
vortices, these results indicate that our model is useful for future
technological applications involving intrinsic LJJ stacks. 

\vskip 0.8cm
\centerline{\bf ACKNOWLEDGMENTS}
\vskip 0.4cm

J.H.K. would like to thank H. J. Lee, W. Schwalm and D. H. Wu for
helpful discussions.  This work was supported in part by the UND
Research Seed Money Program.

\newpage

\centerline {\bf FIGURE CAPTIONS} 

\vspace{0.2in}

\noindent {\bf Figure 1.}
A stack of LJJ is shown schematically as alternating superconducting 
(S) and insulating (I) layers with thickness $d_S$ and $d_I$, 
respectively.  $L_x$ denotes the dimension in $x-$direction.  The
magnetic field $B$ is applied in the plane of tunnel barriers and
the bias current $I_B$ is applied along the vertical stack.

\vspace{0.2in}

\noindent {\bf Figure 2.}
Mutual phase-locking of Josephson vortices (filled ovals) with the
wave number $k_m=m\pi/(N+1)a$ is schematically illustrated.  $N$ is
the number of LJJ in the stack and $a=d_S+d_I$.  For the
{\it in-phase} mode ($m=1$), the Josephson vortices form a
rectangular lattice, but for the {\it out-of-phase} mode ($m=N$), 
they form a triangular lattice.  The dotted lines are a guide to
eyes for phase-locking, and the arrows indicate the direction of
propagation.

\vspace{0.2in}

\noindent {\bf Figure 3.}
The dispersion relation for (a) the longitudinal plasma excitations 
at $k_x=0$ and (b) the transverse plasma excitations at $k_m=0$ are
plotted to illustrate the difference between (a) our model and the
magnetic induction model, and (b) our model and the charging effect
model.  Here $\alpha=0.2$ and ${\cal S}= -0.49999$ are chosen. 

\vspace{0.2in}

\noindent {\bf Figure 4.}
The plasmon dispersion in the Meissner state (i.e.,
$\varphi^{(0)}_{\ell,\ell-1}=0$) is plotted as functions of $k_ma$ 
and $k_x\lambda_c$ for the parameters corresponding to (a)
the Nb-Al/AlO$_x$-Nb multilayers ($\alpha=0.0$, ${\cal S}=-0.47$) and
(b) Bi$_2$Sr$_2$CaCu$_2$O$_{8+y}$ ($\alpha=0.2$, ${\cal S}=-0.49999$).

\vspace{0.2in}

\noindent {\bf Figure 5.}
Stability diagram of Mathieu's equation.  The unstable regions and
the stable regions are shaded and not shaded, respectively.  The
periodic solutions of period $2\pi$ (dashed lines) and $4\pi$ (solid
lines) represent the boundary between the stable and unstable regions. 
The filled squares represent the values of (${\bar\delta}_m^T$,
${\bar\eta}_m^T$) satisfying the stability condition. 

\vspace{0.2in}

\noindent {\bf Figure 6.}
The I-V curves for the supercurrent branch in the resistive state are
plotted for (a) $\alpha=0.0$ and (b) 0.1 to illustrate the effect of
nonzero $\alpha$ (i.e., charging effect).  Here, only three curves
corresponding to the collective modes shown in Fig. 2 are plotted for
clarity.

\end{document}